\documentclass{PoS}

\title{Search for New Physics with Rare Heavy Flavour Decays at LHCb}

\ShortTitle{Rare Decays at LHCb}

\author{\speaker{Giampiero Mancinelli}\thanks{on behalf of the LHCb Collaboration}\\
        CPPM, Aix-Marseille Universit\'e, CNRS/IN2P3, Marseille, France \\
        E-mail: \email{giampi@cppm.in2p3.fr}}

\abstract{The LHCb experiment has the potential, during the 2010-11 run, to observe the rare decay $B^0_s\to \mu^+\mu^-$ or improve significantly its exclusion limits. This study will provide very sensitive probes of New Physics (NP) effects.
High sensitivity to NP contributions is also achieved by measuring photon polarization by performing a time dependent analysis of $B^0_s \to \phi\gamma$, and by an angular study of the decay $B^0_d \to K^{*0}\mu^+\mu^-$.
Preparations for these analyses are presented and studies shown of how existing data, for example prompt $J/\psi$ events, can be used to validate the analysis strategy.}

\FullConference{35th International Conference of High Energy Physics - ICHEP2010,\\
		July 22-28, 2010\\
		Paris France}

\begin{document}

\section{Introduction}

The study of rare $B$ decays is a completely complementary approach with respect to the one from direct searches being carried out at general purpose detectors at the LHC.
In fact, yet undiscovered particles can be produced and observed as real particles, or they can appear as virtual particles in loop processes, whose effects can be observed as deviations from the Standard Model (SM) expectation.
This can be done via CP violation tests or with the help of very rare decays which are the topic of this contribution. In particular we give an overview of the main discovery channels: $B^0_s\to\mu^+\mu^-$, $B^0_d\to K^*\mu^+\mu^-$ and $B^0_s\to\phi\gamma$. 
These are processes involving Flavor Changing Neutral Currents (FCNC). They constitute the ideal environment for indirect searches, as they are suppressed in the SM and only realized via boxes or penguin diagrams. Here the "pollution" from the SM is typically very small and the supposedly small effects from NP can show up at the same, or even higher, level with respect to the SM ones.
Furthermore, were NP to be discovered by CMS or ATLAS, its real structure might only be understood via indirect measurements, which can give access to the phases of the new couplings, hence to the flavour structure of NP.

To study these very rare processes high statistics are needed. This year's goal for the LHC is to reach an instantaneous luminosity of $10^{32}$ {\rm cm}$^{-2}$ {\rm s}$^{-1}$, which is only a factor 2 less than the LHCb nominal working value. The expected integrated luminosity by the end of 2011 is 1 {\rm fb}$^{-1}$. Most of the results presented here use 15 {\rm nb}$^{-1}$ of data.
The LHCb detector is a one arm spectrometer about 20 {\rm m} long covering about 300 {\rm mrad} in polar angle around the beams. It has all the elements of a typical multipurpose detector. It is a detector optimized for $B$ physics, with excellent spatial resolution and vertex separation, needed for time dependent measurements, good particle identification, crucial for trigger and flavor-tagging, and remarkable momentum, hence mass, resolution~\cite{LHCB}. 

\section{Rare Decays}

In the following we present the preparation for some of the key measurements involving rare $B$ decays ongoing with the very first data collected at the LHC. More details can be found in the LHCb roadmap document~\cite{ROADMAP}.

\subsection{Search for $B^0_s\to\mu^+\mu^-$}

Within the SM the dominant contribution for this mode stems from the $Z-$penguin diagram, while the box diagram is suppressed by a factor of $|M_W/m_t|^2$. This is a FCNC mode and is also helicity suppressed, hence the SM expectation is only $3.2$ $10^{-9}$~\cite{BSMUMUEXP}. The uncertainty on the theoretical expectation is small as well, which makes this mode very attractive as a SM test bench. This mode is very sensitive to NP, especially in models with an extended Higgs sector and high $\tan\beta$. For example in the MSSM the $B^0_s\to\mu^+\mu^-$ branching ratio (BR) goes as $(\tan\beta)^6$. 
The current 90\% CL limit from the Tevatron is $36\times 10^{-9}$~\cite{CDF}, hence there is large margin for discovery. With 2 $fb^{-1}$ we expect to cover all the interesting parameter space in the NUHM model~\cite{NUHM}. 

In LHCb, the $\mu^+ \mu^-$ pairs are selected on triggered events with criteria very close to the ones used for the control samples in order to keep systematic uncertainties as low as possible. Each $B^0_s\to\mu^+\mu^-$ candidate is given a likelihood to be signal or background-like in a 3D space formed by the ensemble of geometrical event variables, the invariant mass, and the particle identification (PID) information. The three likelihoods are largely uncorrelated. 
To translate the number of observed candidates into a BR measurement, a normalization channel is needed (as $B^0_{d/s}\to h^+ h'^-$ or $B^-\to J/\psi K^-$), as well as the knowledge of the relative efficiencies. The largest systematic uncertainty (13\%) comes to the uncertainty on the $B_s$ and $B_{d/u}$ hadronization fractions, $f_s$ and $f_{d/u}$. A new method to measure $f_s/f_d$ with $B\to DK$ and $B_s \to D_s\pi$ decays is suggested~\cite{FSFD} and should lower this systematic as well as allow a measurement of $f_s/f_d$ at LHC energies.

\begin{figure}[thb]

  \begin{minipage}{0.475\linewidth}
    \centering
    \mbox{\includegraphics[width=0.93\linewidth]{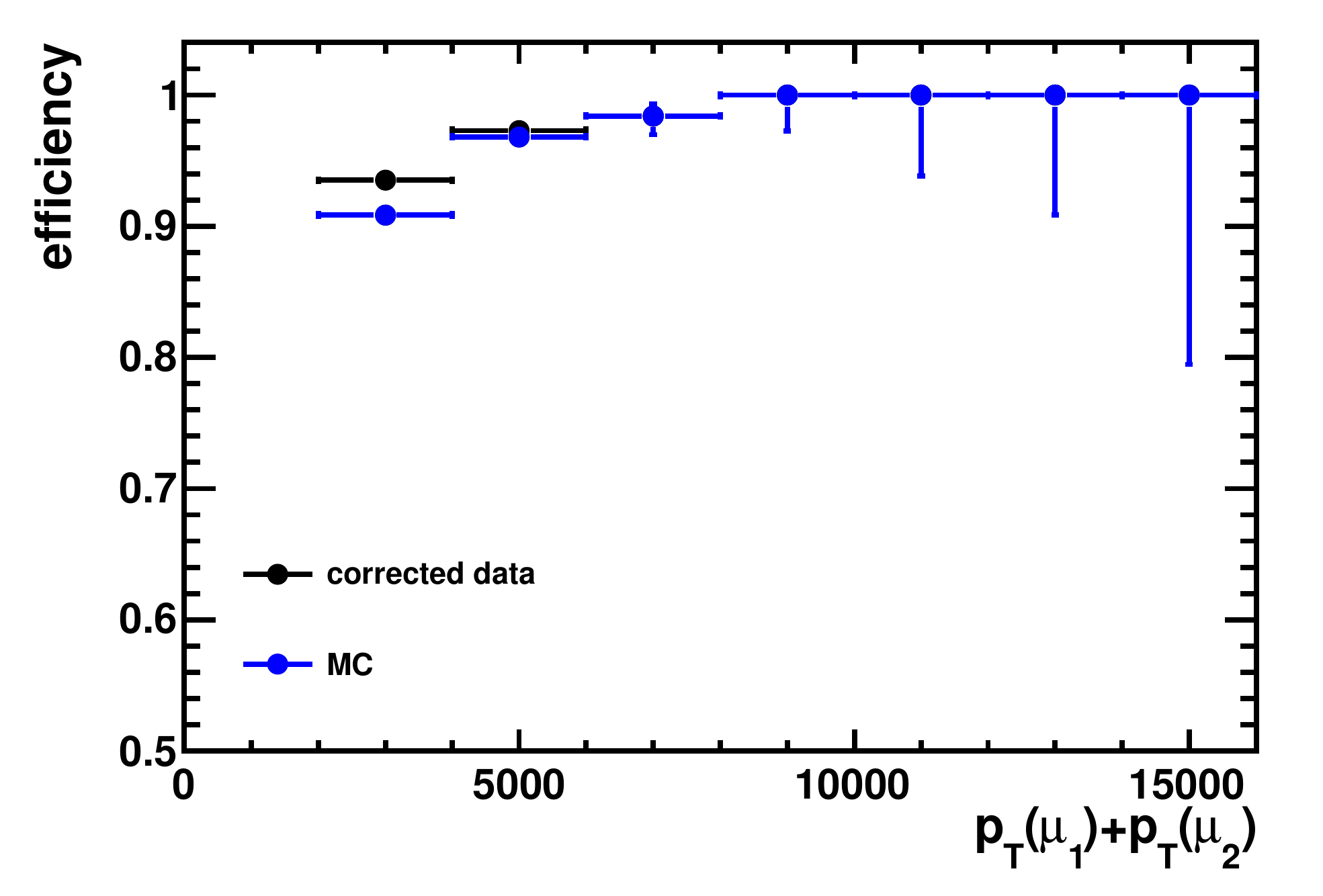}}     
    \caption{Trigger efficiency as a function of the $P_T$ of the $J/\psi$ candidates for data (black) and simulated signal (blue).}
    \label{fig:trig1}
  \end{minipage}
  \begin{minipage}{0.05\linewidth}
  \end{minipage}
  \begin{minipage}{0.475\linewidth}
    \centering
    \mbox{\includegraphics[width=0.93\linewidth]{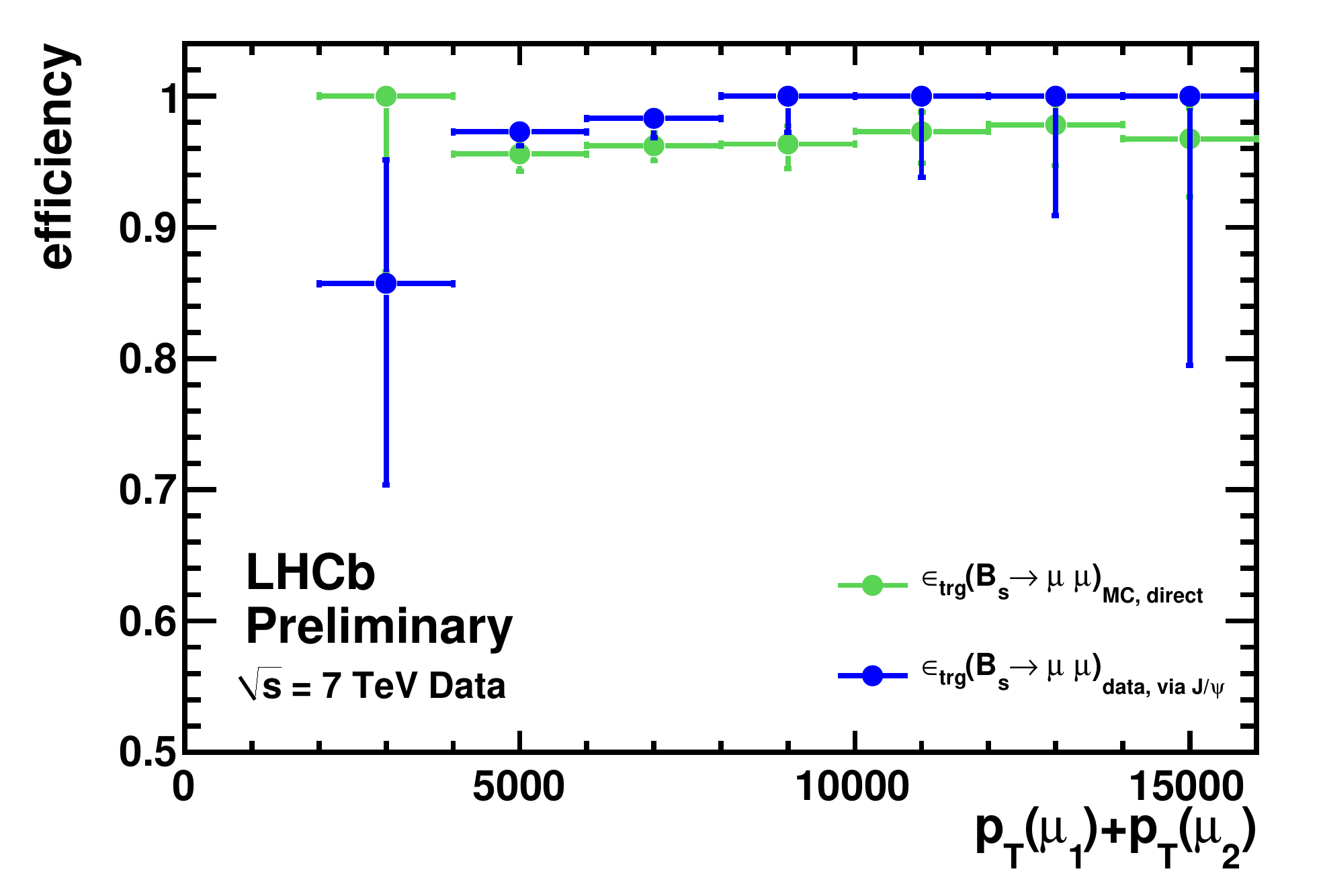}}     
    \caption{Trigger efficiency as a function of the $P_T$ of the $B_s$ for simulation (green) and for data, based on the $J/\psi$ control sample (blue).}
    \label{fig:trig3}
  \end{minipage}
  
\end{figure}

A first example of the use of control samples for the determination of the trigger efficiency is shown in figures~\ref{fig:trig1} and \ref{fig:trig3}. The first shows the trigger efficiency for $J/\psi\to \mu^+ \mu^-$ decays in data and Monte Carlo (MC) simulation, which are in reasonable agreement. With this knowledge we can convolute the efficiencies with the spectrum of the transverse momentum ($P_T$) of the $B_s$ signal from simulation to obtain the expected efficiencies shown in figure~\ref{fig:trig3}. According to the simulation, this procedure is accurate to 1.3\% and the efficiency we can expect from data for $B^0_s\to \mu^+ \mu^-$ is around 99\% with the current trigger settings.
$J/\psi$, $K_s$, and $\Lambda$ decays are used to calculate PID performances. Data and simulation are already in very good agreement. The expected mass resolution is about 20 MeV/c$^2$, though currently the performances on data are not yet as good as what is expected from the simulation in the $J/\psi$ control sample; ultimately $B^0_s\to K^+K^-$ will be used to extract the resolution from the data.

\begin{figure}[thb]

  \begin{minipage}{0.475\linewidth}
    \centering
    \mbox{\includegraphics[width=0.84\linewidth]{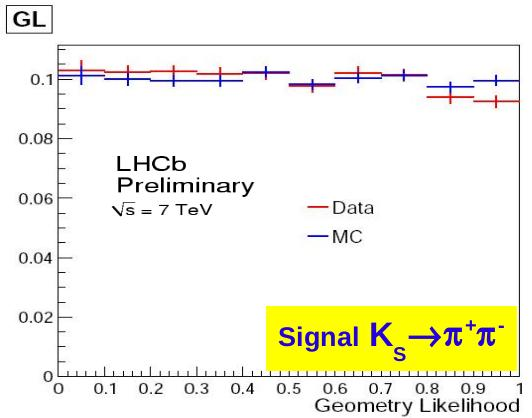}}     
    \caption{GL distribution for the $K^0_s$ control sample for data (red) and simulation (blue).}
    \label{fig:KS0_GL_3var}
  \end{minipage}
  \begin{minipage}{0.05\linewidth}
  \end{minipage}
  \begin{minipage}{0.475\linewidth}
    \centering
    \mbox{\includegraphics[width=0.88\linewidth]{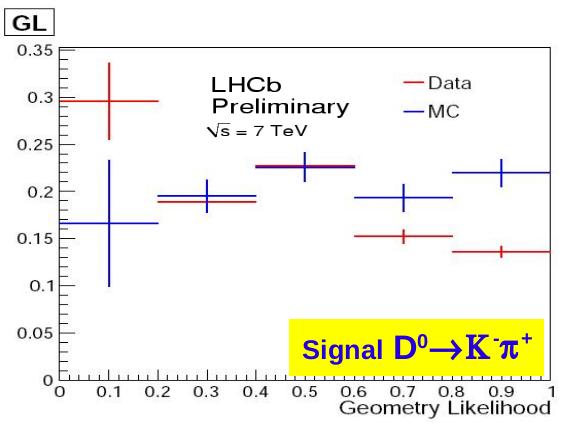}}     
    \caption{GL distribution for the $D^0\to K^-\pi^+$ control sample for data (red) and simulation (blue).}
    \label{fig:D0_GL_4var}
  \end{minipage}
  
\end{figure}

\begin{figure}[thb]

  \begin{minipage}{0.475\linewidth}
    \centering
    \mbox{\includegraphics[width=0.93\linewidth]{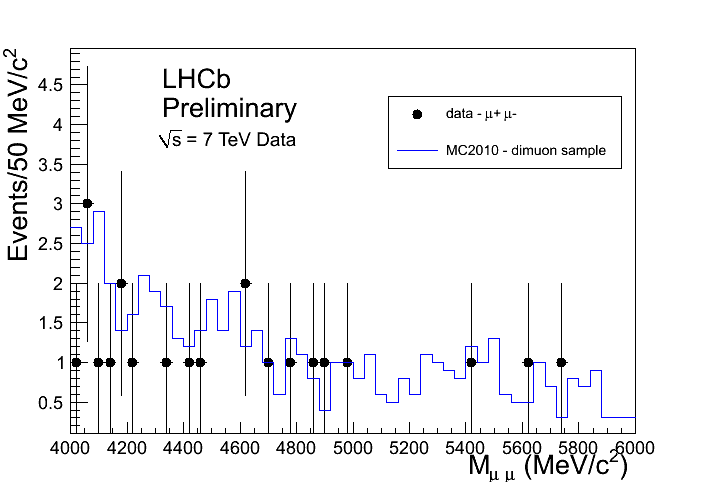}}     
    \caption{$\mu^+\mu^-$ pairs candidates mass distribution for data (dots) and simulation (blue).}
    \label{fig:bsmumu_bkg}
  \end{minipage}
  \begin{minipage}{0.05\linewidth}
  \end{minipage}
  \begin{minipage}{0.475\linewidth}
    \centering
    \mbox{\includegraphics[width=0.91\linewidth]{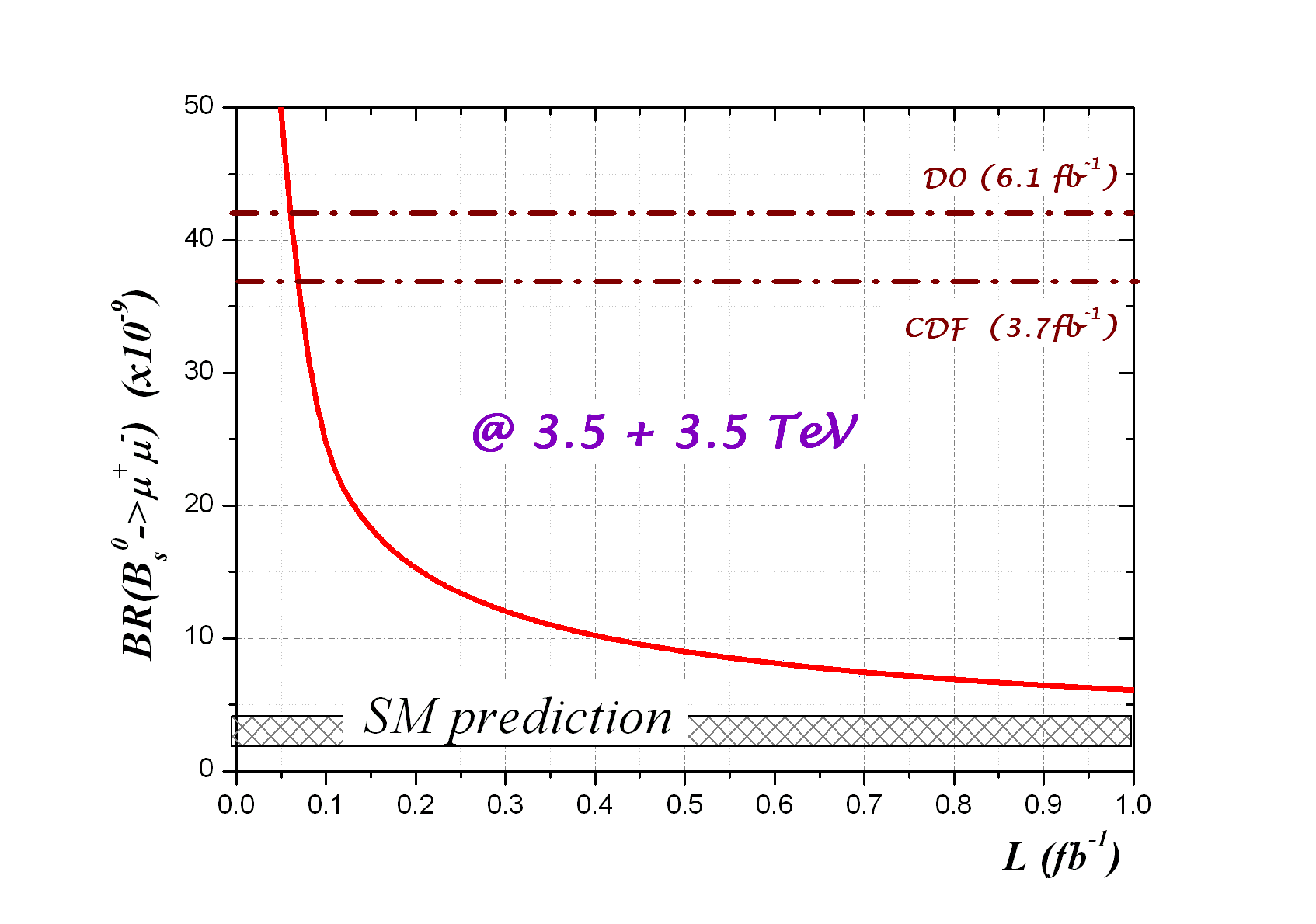}}     
    \caption{Expected exclusion limits at 90$\%$ CL for the $B^0_s\to\mu^+\mu^-$ BR as a function of the luminosity.}
    \label{fig:for_Giampi_3}
  \end{minipage}
  
\end{figure}

The geometrical likelihood (GL) combines information from observables like impact parameters, vertex $\chi^2$, and muon isolation. Developed with the help of simulated data, the signal response will eventually be calibrated with $B\to h h'$ events. For the moment, the machinery is tested on two body decays of long lived lower mass resonances. The distributions of the GL for signal data and simulation for $K_s\to \pi^+\pi^-$ and $D^0 \to K^-\pi^+$ are shown in figures~\ref{fig:KS0_GL_3var} and~\ref{fig:D0_GL_4var}: they are in good but, at the moment, not perfect agreement.
All studies done on data so far confirm the sensitivity calculated on simulated data. Furthermore the background level (see figure~\ref{fig:bsmumu_bkg}) appears to be well simulated, though the dimuon data at high masses is at the moment very limited.

In conclusion, in absence of signal, the current 90\% CL limit from CDF will be improved with less than 100 pb$^{-1}$ of data and with less that 200 pb$^{-1}$ LHCb should reach the final expected Tevatron limit (see figure~\ref{fig:for_Giampi_3}). At the end of next year LHCb should be able to exclude any significant enhancement from the SM at 90\% CL. In presence of a signal at the end of next year LHCb should be able to observe, at 5 $\sigma$, a signal as large as 3.5 the SM expectation, while to reach the SM BR, 10 fb$^{-1}$ at 14 TeV are needed. NP discovery is expected with just 1 fb$^{-1}$ for BR already as small as $17-20 \times 10^{-9}$ depending on how well $f_d/f_s$ can be measured.  

\subsection{Asymmetries in $B^0_d\to K^{*0}\mu^+\mu^-$ decays}

This decay, much less rare (it has a BR about $1.0\times 10^{-6}$ in the SM), is sensitive to magnetic, vector, and axial semileptonic penguin operators, whose coefficients appear in the differential decay rate formula.
Depending on the NP hypotheses, the differential decay rate as a function of the dimuon mass takes a different form, hence, even though the total BR measurements agree with the SM prediction within 20\%, NP contributions can affect other composite variable and asymmetries. In particular, the Forward-Backward Asymmetry ($A_{FB}$) in the $\mu^+\mu^-$ rest frame as a function of the $q^2$ of the muons will be the focus of the first analyses with this mode in LHCb.

What is enticing in this measurement is that at the zero point of $A_{FB}$, the dominant theoretical uncertainties due to form factors calculations cancel out. Assuming that LHCb finds the same central value of $A_{FB}$ at low $q^2$ found by the Belle collaboration~\cite{BELLE}, with 100 pb$^{-1}$ LHCb would achieve an uncertainty similar to those of current best measurements and by the end of next year, having collected about 1400 events, it might reach a 4 $\sigma$ discrepancy with respect to the SM.

\subsection{Photon polarization with $B^0_s\to\phi\gamma$ and $B^0_d \to K^{*0}l^+l^-$ decays} 

Already observed by the Belle Collaboration~\cite{BELLE2} this process $B^0_s\to\phi\gamma$ is even less rare than previous ones. The inclusive results from the BR of $b\to s\gamma$, in good agreement with
the SM expectations, imply stringent constraints on NP contributions. 
The exclusive BR are not very interesting as they suffer from very large theoretical uncertainties. 
The $Q_7$ operator, which is the main operator responsible for $b\to s\gamma$ processes,
produces almost exclusively left-polarized photons in the SM. This
implies that $\phi\gamma$ is not a CP eigenstate. One can in any case consider the fraction of wrongly polarized photons in a "semi-standard" time dependent CP analysis.
In $B_d$, where $\Delta \Gamma \sim 0$, a flavour-tagged
time-dependent analysis is required to obtain sensitivity to the photon
polarization. In $B_s$ instead, where $\Delta \Gamma/\Gamma$ is
significantly different from 0, the flavour averaged decay rate also has
sensitivity to the polarization, making an analysis without flavour
tagging possible.
The energy calibration at LHCb is already promising. Currently it is based on low mass resonances, until a sample of $K^{*0}\gamma$ will be obtained. We expect in one year about 4800 $B^0_s\to \phi\gamma$ events which should give sensitivities of the order or better than the current world averages.

Another way to measure photon polarization is via $B^0_d\to K^{*0}e^+e^-$ decays for very low $e^+e^-$ invariant masses. The available statistics will be smaller and large backgrounds are expected. Nonetheless the systematics are easier and different from the ones of $B^0_s\to\phi\gamma$, hence this mode is expected to be equally powerful.
Finally $B^0_d\to K^{*0}\mu^+\mu^-$ decays for low $\mu^+\mu^-$ masses can also be used. Because of the higher masses, the previous analysis can not be just duplicated and the terms involved are more complex. On the other hand it will be possible to study other interesting observables from the interference.

\section{Conclusions}

First $B$ decays have been recorded in the LHCb detector, which is well on track for its heavy flavour program. The first validation work with 2010 data is very promising: trigger, particle identification, and tracking efficiencies are reasonably close to expectations and the prospects are already interesting with 100 pb$^{-1}$ of data.

\end{document}